# Mapping Charge Excitations in Generalized Wigner Crystals

*Hongyuan Li*[1, 2, 3, 8], *Ziyu Xiang*[1, 2, 3, 8], *Emma Regan*[1, 2, 3], *Wenyu Zhao*[1], *Renee Sailus*[4], *Rounak Banerjee*[4], *Takashi Taniguchi*[5], *Kenji Watanabe*[6], *Sefaattin Tongay*[4], *Alex Zettl*[1, 3, 7], *Michael F. Crommie*[1, 3, 7]\*, and *Feng Wang*[1, 3, 7]\*

[1]Department of Physics, University of California at Berkeley, Berkeley, CA, USA.

[2]Graduate Group in Applied Science and Technology, University of California at Berkeley, Berkeley, CA, USA.

[3]Materials Sciences Division, Lawrence Berkeley National Laboratory, Berkeley, CA, USA.

[4]School for Engineering of Matter, Transport and Energy, Arizona State University, Tempe, AZ, USA.

[5]International Center for Materials Nanoarchitectonics, National Institute for Materials Science, Tsukuba, Japan

[6]Research Center for Functional Materials, National Institute for Materials Science, Tsukuba, Japan

[7]Kavli Energy Nano Sciences Institute at the University of California Berkeley and the Lawrence Berkeley National Laboratory, Berkeley, CA, USA.

[8]These authors contributed equally: Hongyuan Li and Ziyu Xiang

**Abstract:** Transition metal dichalcogenide-based moiré superlattices exhibit very strong electron-electron correlations, thus giving rise to strongly correlated quantum phenomena such as generalized Wigner crystal states[1-10]. Theoretical studies predict that unusual quasiparticle excitations across the correlated gap between upper and lower Hubbard bands can arise due to long-range Coulomb interactions in generalized Wigner crystal states[7,9,11]. Here we describe a new scanning single-electron charging (SSEC) spectroscopy technique with nanometer spatial resolution and single-electron charge resolution that enables us to directly image electron and hole wavefunctions and to determine the thermodynamic gap of generalized Wigner crystal states in twisted $WS_2$ moiré heterostructures. High-resolution SSEC spectroscopy was achieved by combining scanning tunneling microscopy (STM) with a monolayer graphene sensing layer, thus enabling the generation of individual electron and hole quasiparticles in generalized Wigner crystals. We show that electron and hole quasiparticles have complementary wavefunction distributions and that thermodynamic gaps of order 50meV exist for the 1/3 and 2/3 generalized Wigner crystal states.

A Wigner crystal is the crystalline phase of electrons stabilized at low electron density when long-range Coulomb interactions dominate over quantum fluctuations in electron motion[12]. The recent discovery of flat moiré minibands in van der Waals heterostructures has opened a new route to realize Wigner crystal states at zero magnetic field[1-8]. A variety of generalized Wigner crystal states have been reported in transition metal dichalcogenide (TMDC) moiré superlattices[1-5], and real-space imaging of the electron lattice of generalized Wigner crystals has been performed using a new form of non-invasive STM imaging[5]. A microscopic understanding of generalized Wigner crystal excitations, however, is still lacking. Because the electron lattice is fragile, it is challenging to image quasiparticle (e.g., electron/hole) wavefunctions and to spectroscopically determine the correlated gaps of generalized Wigner crystals.

Several scanning probe microscopy techniques have previously been developed to probe fragile correlated states, such as scanning charge accumulation microscopy[13-15] and scanning single-electron transistor microscopy[16-18]. The spatial resolution of these microscopy tools, however, is typically limited to ~100nm, and so is not sufficient to image generalized Wigner crystal quasiparticle states at the single unit cell level. Here we describe a new scanning single-electron charging (SSEC) spectroscopy that has ~1 nm spatial resolution as well as single electron sensitivity. SSEC spectroscopy technique combines STM with a monolayer graphene sensing layer and enables local manipulation of individual electron- and hole-quasiparticles in generalized Wigner crystals via STM tip-based local gating. It enables direct visualization of quasiparticle excitations and spectroscopic determination of the thermodynamic gap of generalized Wigner crystals. Using this technique, we observe that electron and hole quasiparticles excitations exhibit complementary wavefunction distributions and that thermodynamic gaps of order 50meV exist for the 1/3 and 2/3 generalized Wigner crystal states.

Fig. 1a shows the design of the sample and the measurement scheme. The sample is a near-60° twisted WS$_2$ (t-WS$_2$) moiré heterostructure encapsulated in hBN layers. It is dual gated by a monolayer of graphene on top (the top gate) and graphite on the bottom (the bottom gate). The hBN dielectric layer thicknesses for the top and bottom gates are 5.8nm and 37nm, respectively. Sample fabrication details are included in Methods. The charge carrier densities of the t-WS$_2$ and the top graphene sensing layer are tuned independently by applying a bottom gate voltage $V_{BG}$ and a top gate voltage $V_{TG}$. A bias voltage ($V_{bias}$) is applied between the graphene top gate (otherwise known as the sensing layer) and the STM tip. Application of $V_{bias}$ allows electrons in the t-WS$_2$ moiré heterostructure to be manipulated by local tip-gating and to be detected through charging events measured via the tunnel current to the graphene sensing layer.

Conventional STM measurement of the graphene sensing layer provides information on the corrugation of the heterostructure sample as shown in Fig.1b. The thin top graphene and hBN bend conformally and thus inherit the topography of the t-WS$_2$ moiré superlattice. Two sets of moiré superlattices with distinct periods are observed. The larger periodicity (9 nm) originates from the t-WS$_2$ moiré superlattice which has a twist angle of 58°, while the smaller periodicity (~1.5 nm) corresponds to the moiré superlattice formed by the top graphene and hBN which has a twist angle of 9.6°. The 58° t-WS$_2$ moiré superlattice exhibits a triangular lattice with three types of high symmetry stacking regions in each unit cell: a bright region (B$^{S/S}$ stacking), a dark region (AB stacking), and a medium height region (B$^{W/W}$ stacking)[19] (see Fig. 1b inset). The bonding arrangements of the B$^{S/S}$, AB, and B$^{W/W}$ stacking orders are sketched on the left side of Fig. 1b.

Figure 1c-f illustrates the dual role of the STM tip in SSEC spectroscopy. In Fig. 1c the biased STM tip is seen to act as a local gate on the t-WS$_2$ because its electrical field partially

penetrates the monolayer graphene. This is because graphene has a small electron density of states and weak screening, especially when its Fermi level is near the Dirac point. When the sample-tip bias $V_{bias}$ matches the work function difference between the tip and the graphene the tip exerts no local gating effect (Fig. 1d). With a decreased (increased) $V_{bias}$, positive (negative) charge accumulates at the tip apex and generates local band bending in the t-$WS_2$ due to E-field penetrating through the monolayer graphene (Figs. 1e and 1f). With sufficiently strong band bending a single electron (hole) quasiparticle will be injected into the t-$WS_2$. The added charge due to this tip-induced quasiparticle excitation will, in turn, alter the tip-graphene tunnel current via long-range Coulomb interactions. SSEC spectroscopy has some similarity to capacitance spectroscopy in that the tip bias voltage drives charging of the t-$WS_2$. However, unlike conventional capacitance spectroscopy, SSEC spectroscopy locally manipulates individual electrons/holes in the heterostructure and is responsive to individual charge excitation through the tunnel current to the graphene sensing layer. A spatial resolution of ~ 1 nanometer can be achieved in SSEC spectroscopy for thin top hBN layers having a thickness of several nanometers.

Figure 2a shows the backgate voltage ($V_{BG}$) dependence of the dI/dV spectra of the tip-graphene tunnel junction when the tip is placed over the $B^{S/S}$ site of the moiré unit cell. We started by setting the top gate voltage to $V_{TG}$ = 0.52V, which shifts the t-$WS_2$ chemical potential up to the conduction band edge while keeping the graphene Fermi level close to the Dirac point[5]. Under these conditions increasing the backgate voltage, $V_{BG}$, increases the global electron doping in the t-$WS_2$ layers. The resulting dI/dV measurement of the sensing layer is dominated by a broad increase in the dI/dV signal for increased $V_{bias}$ regardless of polarity, which mostly reflects the local density of states of graphene and does not show an obvious dependence on $V_{BG}$ (i.e. on the t-WS$_2$ doping). The impact of electrical charge added to the t-WS$_2$ moiré superlattice is better seen by normalizing the dI/dV spectra at each V$_{BG}$ by the averaged dI/dV spectrum over all V$_{BG}$ values as seen in Fig. 2b (see the SI for normalization details). This normalization removes the broad rising background and reveals multiple dispersive bright lines that correspond to peaks in dI/dV that shift in energy with applied V$_{BG}$. These peaks are clustered around several electron doping levels in the t-WS$_2$ moiré superlattice (i.e., different V$_{BG}$ values), as denoted by horizontal arrows in Fig. 2b. Their V$_{BG}$ values correspond to t-WS$_2$ electron filling factors of n = 0, 1/3, 2/3, 1 (as labeled in red), where n is the number of electrons per moiré site. The filling factors shown here are based on carrier densities extracted using the device capacitance as described in reference[5]. Our SSEC imaging (see SI and Fig.3) is also consistent with these filling factors.

The dI/dV spectra of Fig. 2b show two or more dispersive lines clustered around each correlated insulating state at n = 1/3, 2/3, and 1. To better understand this behavior Fig. 2c shows higher resolution V$_{BG}$-dependent dI/dV spectra near the n = 2/3 generalized Wigner crystal state (the phase space inside the white dashed box in Fig. 2b). Two bright dispersive lines with similar slope move together through the V$_{BG}$ region associated with the n = 2/3 state, as well as several weaker features nearby. Figure 2d displays a horizontal line cut of Fig. 2c at V$_{BG}$ = 1.60V, where clear dI/dV peaks (labeled with vertical arrows) can be observed at the V$_{bias}$ positions of the bright lines in Fig. 2c. These dI/dV peaks do not mark the energy locations of resonances in the local density of states (LDOS), but rather arise from electron and hole charging events in the generalized Wigner crystal states of t-WS$_2$.

To understand this, we look to the sketch in Fig. 2e that illustrates the charging diagram of the t-WS$_2$ moiré superlattice in the n = 2/3 state. There are three different regimes as shown in

Fig. 2f: the "intrinsic" generalized Wigner crystal insulator phase (I) (solid red dots mark the locations of electrons in the moiré unit cell while open circles mark empty cells), the electron excitation regime (E) where an electron (blue solid dot in Fig. 2f) is injected below the tip at a large negative $V_{bias}$, and the hole excitation regime (H) where a hole (blue open circle in Fig. 2f) is injected at a large positive $V_{bias}$. These regimes are separated by two dispersive lines in the $V_{BG}$-$V_{bias}$ parameter space in Fig. 2e with the slope of the dispersive lines being determined by the efficiency of the tip as an effective top gate relative to chemical potential shifts induced by the bottom gate. Starting from the intrinsic regime (e.g., Fig. 1d), a reduction of $V_{bias}$ causes the tip to become positively charged. Crossing the boundary from (I) to (E) corresponds to pushing the UHB energetically below the t-WS$_2$ chemical potential $\mu_{tWS_2}$ and locally inducing an electron charging event (Fig. 1e). Similarly, if $V_{bias}$ is increased and crosses the boundary from (I) to (H) then the LHB is energetically pushed above $\mu_{tWS_2}$, resulting in a local hole charging event (Fig. 1f). These charge excitations alter the tunnel junction conductance and result in a peak in the dI/dV spectra. The dispersive dI/dV peaks in Figs. 2b,c correspond to the electron/hole excitation boundaries sketched in Fig. 2e. The reason that the intrinsic region does not bracket $V_{bias} = 0$ is most likely because of the work function difference between the tip and the graphene sensing layer. Additional weaker dI/dV peak features observed in Fig. 2c at higher positive (negative) $V_{bias}$ correspond to the injection of additional electrons and holes at nearby moiré cells adjacent to the tip location.

To establish our assignment of the features seen in Fig.2 as electron and hole excitations of the generalized Wigner crystal, we directly image these charging events using SSEC spectroscopy. Fig. 3a displays an STM topography image of a highly defect-free t-WS$_2$ moiré region chosen for imaging electron/hole excitations. Fig. 3b shows a dI/dV map measured over this same area at $V_{BG}$ = 1.50V, $V_{TG}$ = 0.52V, and $V_{bias}$ = -0.59V, which corresponds to the electron excitation boundary denoted by the filled circle in Fig. 2c. A triangular lattice of bright dots is seen with a period larger than the underlying moiré superlattice by a factor of $\sqrt{3}$. This new triangular lattice reflects the wavefunction distribution of the excited electron in the generalized n = 2/3 Wigner crystal. To confirm that this pattern originates from tip-induced electron excitations we tested how it evolves with $V_{bias}$. A typical aspect of charging features is "ring expansion" with increased bias[20-24] because the tip can then induce charging events from more distant positions. Figs. 3d-g show the evolution of the charging rings with increasing $|V_{bias}|$, obtained at $V_{BG}$ = 1.50V and $V_{TG}$ = 0.52V. The electron charging signal at the different moiré unit cells is seen to expand into a wide charging ring with increasingly negative $V_{bias}$, precisely as expected for electron injection.

Fig. 3c shows a dI/dV map taken at the hole excitation boundary for n = 2/3 filling, corresponding to the open circle in Fig. 2c ($V_{BG}$ = 1.65V, $V_{TG}$ = 0.52V, and $V_{bias}$ = -0.14V). A honeycomb lattice of bright dots (open circles) here reflects the wavefunction of hole excitations in this generalized Wigner crystal. The tip bias dependence of this pattern also confirms its origin as shown in Figs. 3h-k. Here the hole charging signal in the different moiré unit cells is seen to expand into a wider charging ring for increasingly positive $V_{bias}$.

The complementarity of the electron and hole excitation wavefunctions of the generalized Wigner crystal state can be seen by overlaying the hollow circles and solids dots of Figs. 3b,c onto the topography of Fig. 3a. Both the hollow *and* solid circles are seen to be localized in the $B^{W/W}$ stacking regions, where the lowest-energy conduction moiré flat bands are predicted to reside[19]. The electron excitation distribution (red dots) and the hole excitation distribution (open circles) combine perfectly to yield the full moiré superlattice. Hole excitations correspond precisely to the filled electron locations for an n = 2/3 generalized Wigner crystal (i.e., a honeycomb lattice) whereas electron excitations occur at the hollow centers of the honeycomb lattice. This is the pattern that one might intuitively expect from classical electrostatic reasoning.

Since the thermodynamic gap of a correlated state is the chemical potential difference for adding a single hole or electron, it is possible to extract the thermodynamic gap of generalized Wigner crystals from our SSEC spectra. To see this we define the energy difference between the chemical potential and LHB as $\Delta_h$ (Fig. 1d), and the bias applied to the tip to create a hole excitation as $V_h$. We then write $\Delta_h = \alpha_h e V_h$, where $\alpha_h$ is a geometric constant defined by the tip-gating efficiency when the tip is above a hole site (e is the charge of an electron). Similarly, we can write $\Delta_e = \alpha_e(-e)V_e$ where $\Delta_e$ is the energy difference between the chemical potential and UHB, and the factors $V_e$ and $\alpha_e$ are defined for electron excitations. If $\alpha_e = \alpha_h = \alpha$, then the thermodynamic gap, $\Delta = \Delta_e + \Delta_h$, can be written as

$$\Delta = \alpha e(V_h - V_e) = \alpha e \Delta V_{bias}, \qquad (1)$$

where $\Delta V_{bias}$ is the experimental sample-tip bias difference measured between the hole excitation and electron excitation boundaries as shown in Figs. 2c-e. A key requirement in this analysis is that the capacitive coupling between the tip and surface is equivalent for electron and hole excitations (i.e., $\alpha_e = \alpha_h$). This requirement is satisfied in the measurements shown in Fig. 2 which were performed with the tip positioned above the $B^{S/S}$ site in the t-WS$_2$ moiré unit cell, which is the same distance to the nearest excited electron or hole.

In order to obtain a quantitative value of the generalized Wigner crystal thermodynamic gap, $\Delta$, we must determine the value of $\alpha$. This is obtained through numerical simulation of the tip-surface electrostatics. Here we model the tip and the backgate as a metallic cone and a metallic plate, respectively. The screening by Dirac electrons in the graphene sensing layer is included in the simulation (see SI for details). The t-WS$_2$ layer is regarded as a thin insulator since it is in a correlated insulating phase. To obtain $\alpha$, we monitor the local electric potential change, $\Delta\Phi$, in the t-WS$_2$ layer that is induced by applying nonzero V$_{bias}$. The two main variable parameters in the simulation are the metallic cone angle of the tip ($\theta$) and the tip height ($h$). It is also useful to introduce the backgate coupling parameter $\beta$ such that the electric potential change at the t-WS$_2$ moiré site is $\Delta\Phi = e\alpha V_{bias} - e\beta V_{BG}$. The ratio, $\alpha/\beta$, is an experimentally accessible quantity since it is the slope of the dispersive features observed in Figs. 2b,c (charging occurs when the $\Phi$-dependent UHB or LHB, becomes equal to the chemical potential). We are thus able to test the accuracy of our simulation against the experimentally measured ratio $\frac{\alpha}{\beta} = 0.51 \pm 0.01$.

To obtain $\alpha$ we first find the values of $\theta$ and $h$ that best reproduce the experimentally observed ratio of $\alpha/\beta$. Varying $\theta$ and $h$ in the simulation does not significantly affect the values of $\alpha$ and $\beta$ as long as the ratio $\alpha/\beta$ remains unchanged (see Fig. S3). The simulation is thus not sensitive to $\theta$ and $h$, and yield a robust value of $\alpha = 0.16 \pm 0.02$. From Eq. (1) this results in the following experimental thermodynamic gaps for the n = 1/3, 2/3, and 1 correlated states: $\Delta_{n=1/3} = 52 \pm 8$meV, $\Delta_{n=2/3} = 47 \pm 8$meV, and $\Delta_{n=1} = 107 \pm 16$meV (the uncertainty here is calculated from both the standard deviation in our measurement of $\Delta V_{bias}$ and the uncertainty in the simulation of $\alpha$).

In conclusion, we have demonstrated a non-invasive high-resolution microscopic tool that enables us to induce electron/hole excitations locally in 2D generalized Wigner crystal systems. This technique allows us to measure the thermodynamic gaps of generalized Wigner crystals having different filling factors and to map their electron and hole excitations. This technique should be applicable to the characterization of other fragile correlated electron systems.

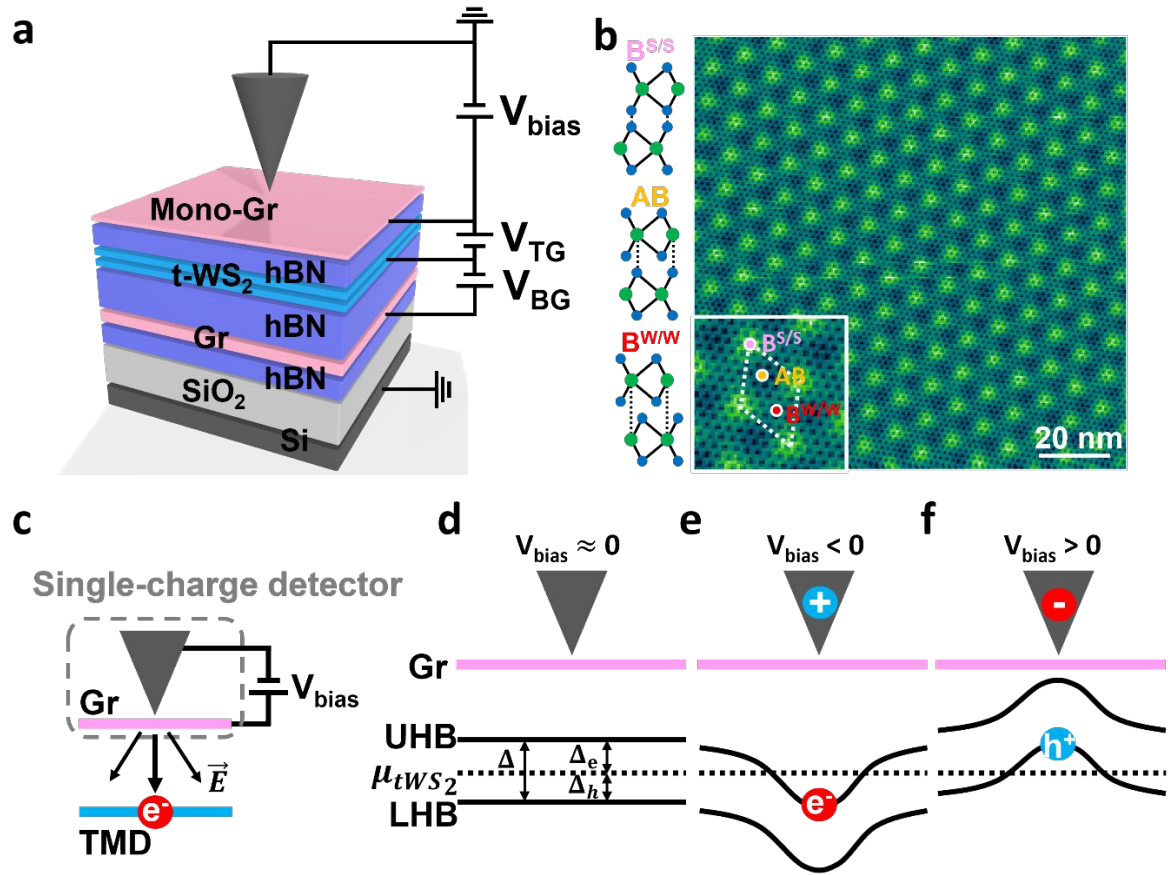

**Figure 1. Scanning single-electron charge spectroscopy measurement of a twisted bilayer WS$_2$ moiré superlattice. a**. Schematic of the dual-gated near-60° twisted bilayer WS$_2$ (t-WS$_2$) moiré superlattice device. The top hBN thickness (5nm) is slightly smaller than the moiré lattice constant (9nm). Top gate ($V_{TG}$) and bottom gate ($V_{BG}$) voltages are separately applied to independently control the carrier densities in the t-WS$_2$ superlattice and top graphene sensing layer. **b**. A typical large-scale topography image of the top graphene ($V_{bias}$ = -0.62V and I = 150

pA). Three different stacking regions are labeled in the close-up image in the inset: $B^{S/S}$ stacking (pink dots), AB stacking (yellow dots), and $B^{W/W}$ (red dots). The structures of the $B^{S/S}$, AB, and $B^{W/W}$ stacking are illustrated in the left pannel **c**. Illustration of scanning single-electron charge (SSEC) spectroscopy. The electric field from the tip bias partially penetrates the graphene and induces quasiparticle excitations in the t-WS$_2$. The tip-graphene tunnel junction detects changes in the number of electrons in the t-WS$_2$ due to long range Coulomb interactions. **d-f**. Illustration of tip-induced quasiparticle excitation in a correlated insulator. Solid curves represent the lower Hubbard band (LHB) and upper Hubbard band (UHB), while dashed line represents the chemical potential $\mu_{tWS_2}$. (**d**) For $V_{bias} \approx 0$, the LHB and UHB are uniform. (**e**) For large $V_{bias} < 0$ the UHB is pushed beneath $E_F$ by tip-gating, inducing a local electron excitation. (**f**) For large $V_{bias} > 0$ tip-gating induces a local hole excitation. For simplicity, we have neglected the work function difference between the tip and the sensing layer graphene in (**d-f**) which may induce a small offset for $V_{bias}$. The gap between the UHB and LHB is labeled as $\Delta$ while the gap between $\mu_{tWS_2}$ and UHB (LHB) is labeled as $\Delta_e$ ($\Delta_h$).

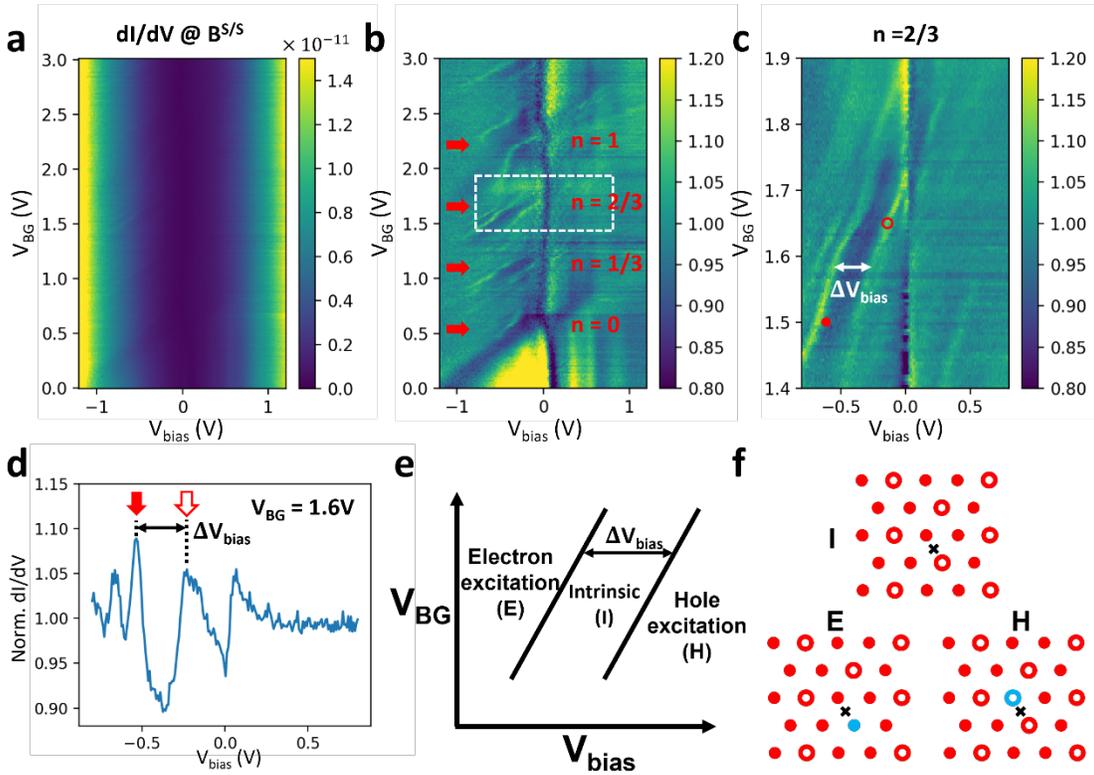

**Figure 2. STS study of quasiparticle excitations in generalized Wigner crystals. a**. dI/dV spectra of the graphene sensing layer as a function of sample-tip bias ($V_{bias}$) and backgate voltage ($V_{BG}$) (measured with tip held over the t-WS$_2$ $B^{S/S}$ site). $V_{TG}$ is fixed at 0.52V. **b**. Normalized form of the dI/dV spectra shown in **a**. The dI/dV spectrum at each $V_{BG}$ is normalized by the average dI/dV spectrum for all $V_{BG}$ values (see SI for details). Dispersive bright lines

corresponding to dI/dV peaks are clustered around the filling factors n = 0, 1/3, 2/3, 1 (labeled in red). **c**. High-resolution dI/dV spectra corresponding to the n = 2/3 state measured over the range enclosed by the white box in **b**. Two bright dispersive lines with similar slopes exist on opposite sides of the n=2/3 state and correspond to electron charging events (solid dot) and hole charging events (open circle), The $\Delta V_{bias}$ offset between them is marked in white. **d**. dI/dV line-cut of **c** at $V_{BG}$ = 1.60V shows peaks corresponding to dispersive features in **c** labeled with a solid arrow (electron charging) and an open arrow (hole charging) that are offset from each other by $\Delta V_{bias}$. **e**. Schematic shows the charging regimes of the t-WS$_2$ moiré superlattice near the n = 2/3 state. (I) marks the intrinsic generalized Wigner crystal insulator phase, (**e**) marks the electron excitation regime, and (H) marks the hole excitation regime. The charging regimes are separated by two dispersive lines in the $V_{BG}$-$V_{bias}$ parameter space that are offset from each other by $\Delta V_{bias}$. **f**. Real space sketch of the intrinsic (I), electron excitation (E), and hole excitation (H) regimes. Electron-filled sites (solid dots) and empty sites (open circles) of the n = 2/3 generalized Wigner crystal as shown. Tip-induced electron excitation is marked by a solid blue dot and hole excitation by a blue open circle. The back cross labels the tip position.

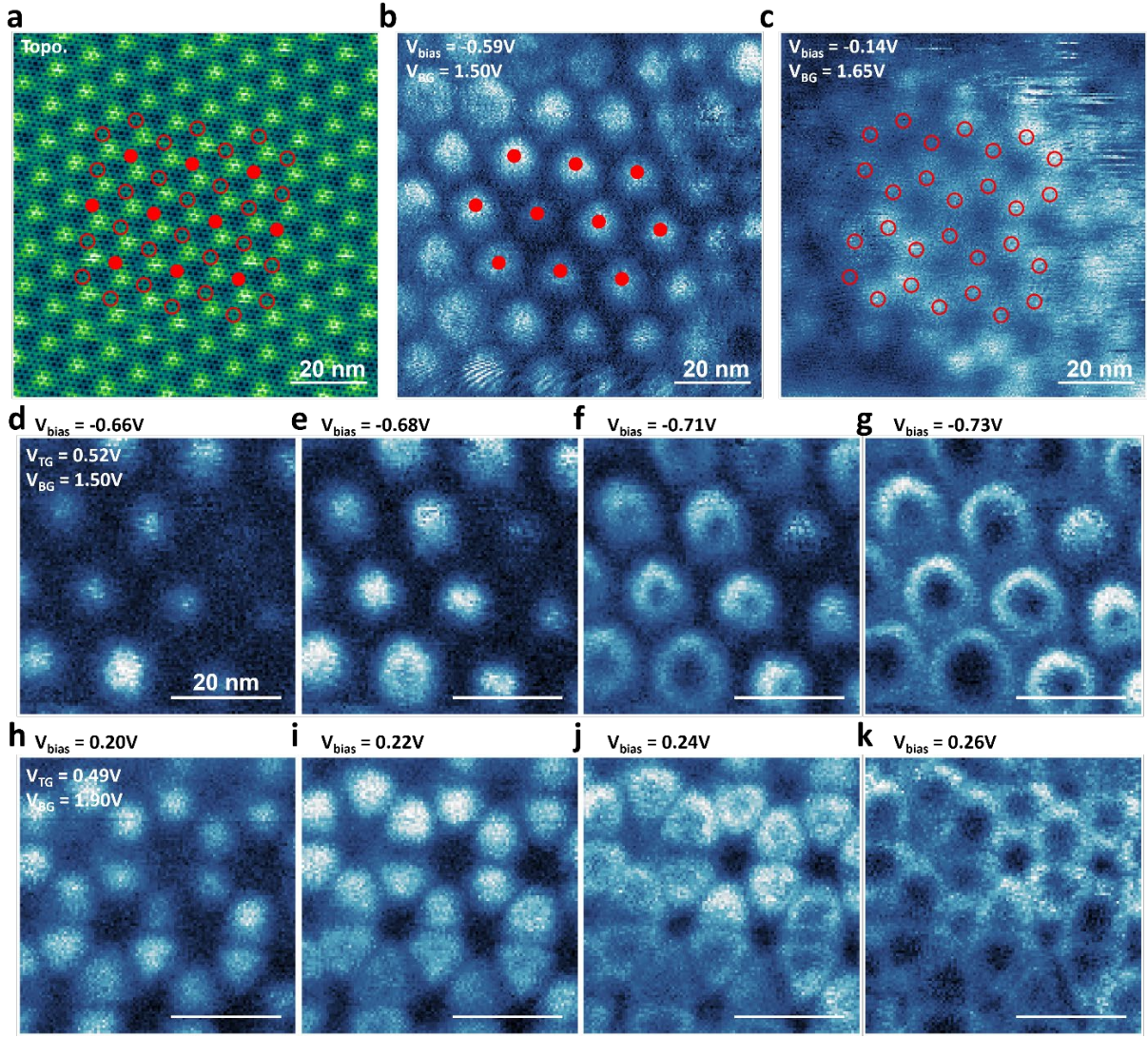

**Figure 3. Mapping electron and hole excitations of the n = 2/3 generalized Wigner crystal.**
**a**. STM topography image of graphene sensing layer shows the t-WS2 moiré superlattice ($V_{bias}$ = -0.59V, I = 150pA). **b**. dI/dV map of same area as (**a**) for applied voltages corresponding to the electron excitation boundary ($V_{BG}$ = 1.50V, $V_{TG}$ = 0.52V, $V_{bias}$ = -0.59V). Sites of electron excitations are marked with solid red dots. **c**. dI/dV map of same area as (**a**) for voltages corresponding to the hole excitation boundary ($V_{BG}$ = 1.65V, $V_{TG}$ = 0.52V, $V_{bias}$ = -0.14V). Sites of hole excitations are marked with open circles. **d-g**. Evolution of dI/dV maps of the electron charging peak of the n = 2/3 state with increasingly negative $V_{bias}$. The electron charging signals widen into a growing circle at each moiré site as $V_{bias}$ becomes more negative. **h-k**. Evolution of dI/dV maps of the hole charging peak of the n = 2/3 state with increasingly positive $V_{bias}$. The hole charging signal widens into a growing circle as $V_{bias}$ becomes more positive. The solid dots in **b** and open circles in **c** are overlaid in **a** and are seen to be perfectly complementary. (**d-k**) share the same scale bar.

**Corresponding Author**

* Email: crommie@physics.berkeley.edu (M.C.) and fengwang76@berkeley.edu (F.W.).

**Author Contributions**

H. L., M.C., and F.W. conceived the project. H.L. and Z.X. performed the STM measurement, H.L., Z.X., E. R., and W.Z. fabricated the heterostructure device. R.S., R.B. and S.T. grew the $WS_2$ crystals. K.W. and T.T. grew the hBN single crystal. All authors discussed the results and wrote the manuscript.

**Notes**

The authors declare no financial competing interests.

**ACKNOWLEDGMENT**

This work was primarily funded by the U.S. Department of Energy, Office of Science, Office of Basic Energy Sciences, Materials Sciences and Engineering Division under Contract No. DE-AC02-05-CH11231 (van der Waals heterostructure program KCFW16) (device electrode preparation and STM spectroscopy). Support was also provided by the US Army Research Office under MURI award W911NF-17-1-0312 (device layer transfer), and by the National Science Foundation Award DMR-1807233 (surface preparation). S.T acknowledges support from DOE-SC0020653 (materials synthesis), NSF DMR-1955889 (magnetic measurements), NSF CMMI-1933214, NSF 2206987, NSF ECCS 2052527, DMR 2111812, and CMMI 2129412. K.W. and T.T. acknowledge support from the Elemental Strategy Initiative conducted by the MEXT, Japan,

Grant Number JPMXP0112101001, JSPS KAKENHI Grant Number JP20H00354 and the CREST(JPMJCR15F3), JST for bulk hBN crystal growth and analysis.

**Methods**

**Sample fabrication:** The encapsulated near-60° twisted $WS_2$ (t-$WS_2$) moire heterostructure stack was fabricated using a micro-mechanical stacking technique[25]. A poly(propylene) carbonate (PPC) film stamp was used to pick up all exfoliated 2D material flakes. The 2D material layers in the main heterostructure region were picked up in the following order: substrate hBN, graphite, bottom hBN, first $WS_2$ monolayer, second monolayer, graphene nanoribbon array (not shown in Fig. 1), top hBN, and then monolayer graphene. The graphene nanoribbon array serves as a contact electrode for the t-$WS_2$ and is fabricated by an electrode-free local anodic oxidation (EFLAO) lithography technique[26]. The two $WS_2$ monolayers are obtained by cutting an originally complete single flake into two pieces using EFLAO to precisely control their crystal directions. The PPC film together with the stacked sample was then peeled, flipped over, and transferred onto a Si/$SiO_2$ substrate ($SiO_2$ thickness 285nm). The PPC layer was subsequently removed using ultrahigh vacuum annealing at 230 °C, resulting in an atomically clean heterostructure suitable for STM measurements. A 50nm Au and 5nm Cr metal layer was evaporated through a shadow mask to form electrical contacts to the graphene layers.

**STS measurement:** A modulation of 25mV amplitude and 500~900 Hz frequency was applied to the tip bias to obtain the dI/dV signal.

**Supplementary Materials**

**Data availability**

The data supporting the findings of this study are included in the main text and in the Supplementary Information files, and are also available from the corresponding authors upon request.

# Supplementary Information for
# Mapping Charge Excitations in Generalized Wigner Crystals

*Hongyuan Li*[1, 2, 3, 8], *Ziyu Xiang*[1, 2, 3, 8], *Emma Regan*[1, 2, 3], *Wenyu Zhao*[1], *Renee Sailus*[4], *Rounak Banerjee*[4], *Takashi Taniguchi*[5], *Kenji Watanabe*[6], *Sefaattin Tongay*[4], *Alex Zettl*[1, 3, 7], *Michael F. Crommie*[1, 3, 7]\*, and *Feng Wang*[1, 3, 7]\*

[1]Department of Physics, University of California at Berkeley, Berkeley, CA, USA.

[2]Graduate Group in Applied Science and Technology, University of California at Berkeley, Berkeley, CA, USA.

[3]Materials Sciences Division, Lawrence Berkeley National Laboratory, Berkeley, CA, USA.

[4]School for Engineering of Matter, Transport and Energy, Arizona State University, Tempe, AZ, USA.

[5]International Center for Materials Nanoarchitectonics, National Institute for Materials Science, Tsukuba, Japan

[6]Research Center for Functional Materials, National Institute for Materials Science, Tsukuba, Japan

[7]Kavli Energy Nano Sciences Institute at the University of California Berkeley and the Lawrence Berkeley National Laboratory, Berkeley, CA, USA.

[8]These authors contributed equally: Hongyuan Li and Ziyu Xiang

1. **Normalization of the dI/dV spectra**

2. **Determination of the filling factors via dI/dV mapping**

3. **Simulation of the tip coupling constant $\alpha$**

1. **Normalization of the dI/dV spectra**

To make charging peaks features in the dI/dV spectra more visible, we reduce the background signal through a special normalization process. Here we take Fig. 2a and 2b in the main text as an example to illustrate this process. We first note that the strong background signal is mainly determined by the graphene density of states, which do not significantly change with the tuning of the backgate voltage $V_{BG}$. On the other hand, the tip-induced charging peaks have a strong dependence on $V_{BG}$ (see detailed explanation in the main text), and show a very dispersive feature in the 2D plot of the $V_{BG}$-dependent dI/dV spectra. To obtain the $V_{BG}$-independent background signal we average all the dI/dV spectra (which average out the fast-changing charging peaks) (blue curve in Fig. S1b). Dividing the raw dI/dV spectra (several typical raw dI/dV spectra are shown in Fig. S1b as red curves) by this averaged dI/dV spectra allows us to obtain the normalized dI/dV spectra. Several typical normalized dI/dV spectra are shown in Fig. S1c. The complete data set of the normalized dI/dV spectra is shown in the 2D color plot in Fig. S1d (same as Fig. 2b in the main text).

The same normalization process is applied to the high-resolution $V_{BG}$-dependent dI/dV spectra measured around the n = 2/3 state shown in Fig. 2c in the main text (with the raw data shown in Fig. S1e). Several typical raw dI/dV spectra and the corresponding normalized spectra are shown in Fig. S1f and S1g, respectively. The complete data set of the normalized dI/dV spectra are shown in Fig. 2c in the main text (also reproduced in Fig. S1h).

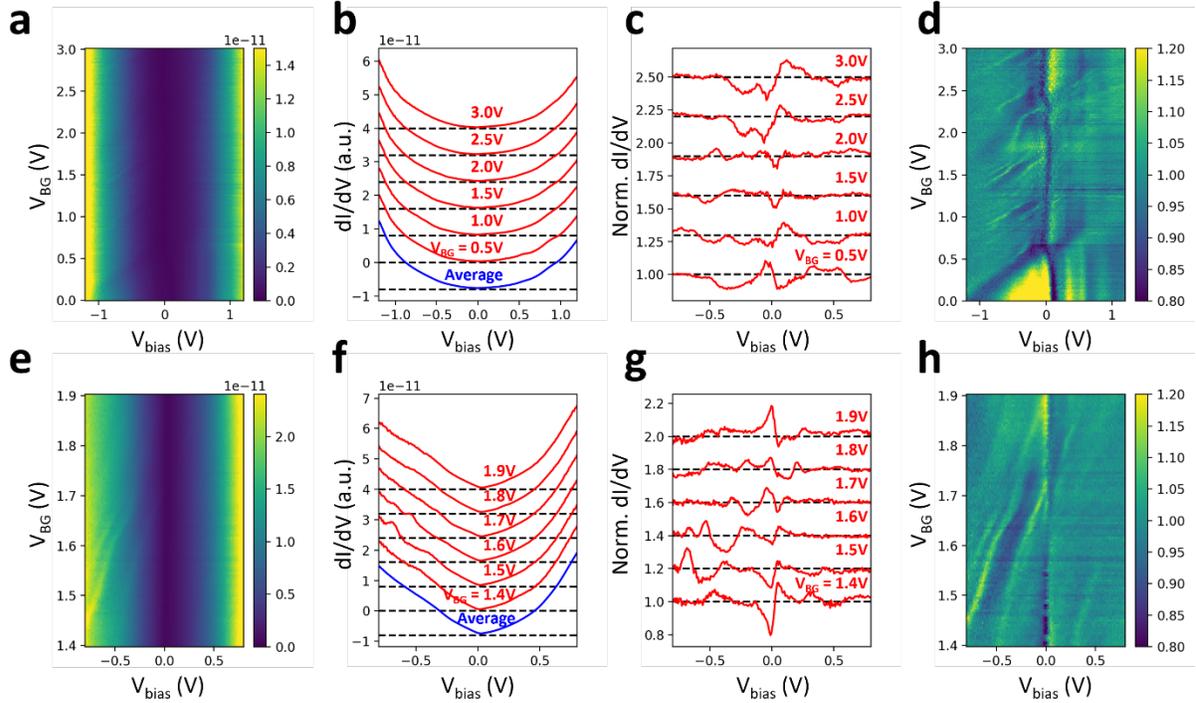

**Figure S1. Normalization of the $V_{BG}$-dependent dI/dV spectra. a**. 2D plot of the raw data for the $V_{BG}$-dependent dI/dV spectra, same as Fig. 2a in the main text. **b**. The red curves show typical dI/dV spectra measured at different $V_{BG}$, while the blue curve shows the mean dI/dV spectra averaged over all spectra shown in **a**. **c**. Normalized dI/dV spectra obtained by dividing the raw spectra (red curves) by the averaged spectra (blue curves) shown in **b**. **d**. 2D plot of the normalized data shown in **a**, same as Fig. 2b in the main text. **e**. 2D plot of the raw data for the dI/dV spectra shown in Fig. 2c of the main text. **f**. The red curves show several typical dI/dV spectra measured at different $V_{BG}$, while the blue curve shows the mean dI/dV spectra averaged over all spectra shown in **e**. **g**. Normalized dI/dV spectra obtained by dividing the raw spectra (red curves) by the average spectra (blue curve) shown in **f**. **g**. 2D plot of the normalized data shown in **e**, same as Fig. 2c in the main text. **h**. 2D plot of the normalized data shown in **e**.

Spectra in **b**, **c**, **f**, and **g** are shifted vertically for clarity, with the corresponding zero reference point labeled by black dashed lines.

## 2. Determination of the filling factors via dI/dV mapping

The filling factor of the moire superlattice is not only determined by the simple capacitor model, but also directly through dI/dV mapping of the electron lattice. For example, the filling factor of the n=2/3 state can be determined through the mapping shown in Fig. 3 in the main text. Several typical results are shown in Fig. S2, including mapping for addition of electrons to the n = 0 state (Fig. S2a and S2b) and addition of holes in the n = 1 state (Fig. S2c and S2d).

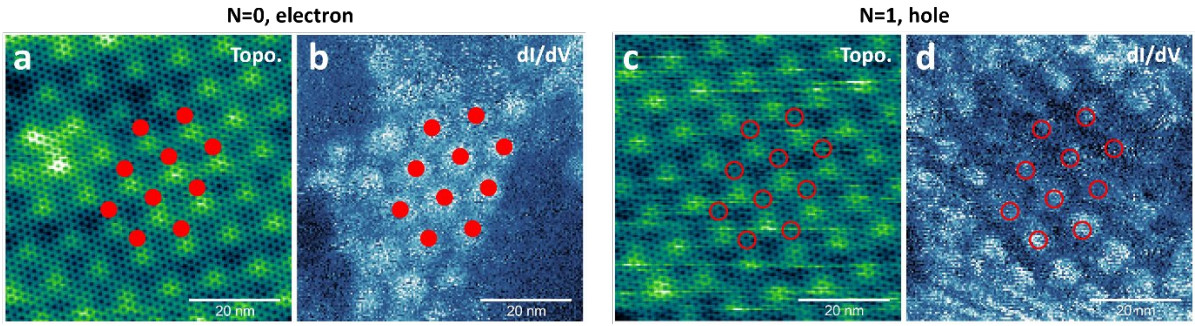

**Figure S2. Determination of the filling factor. a.** Typical STM topography of sensing layer. Red dots denote electron excitation locations. **b**. dI/dV mapping of the electron quasiparticle in the n = 0 state measured in the same area as in **a** ($V_{bias}$ = -0.75V, $V_{BG}$ = 0.43V, $V_{TG}$ = 0.52V). The position of the electrons is labeled with red solid dots **c**. Typical STM topography of sensing layer. Red circles denote hole excitation locations. **d**. dI/dV mapping of the hole quasiparticle in the n =1 state measured in the same area as in **c** ($V_{bias}$ = 0.20V, $V_{BG}$ = 2.4V, $V_{TG}$ = 0.52V). The positions of the holes are labeled with red circles.

3. **Simulation of the tip coupling constant $\alpha$**

In this section, we introduce details for the electrostatic simulation of the tip-t-WS$_2$ coupling constant $\alpha$, including a graphene quantum capacitance effect. As illustrated in Fig. S3a, the tip is approximated by be an ideal metallic cone with a half cone angle $\theta$ and tip height $h$ (separation between tip apex and the graphene surface). The backgate graphite is modeled by an infinitely large metallic plate. The t-WS$_2$ moire heterostructure is regarded as a thin insulator since it is in a correlated insulating state. The t-WS$_2$ is incorporated into the surrounding hBN and regarded as an insulator with the same dielectric constant as the hBN. The graphene is modeled by a special boundary condition whose electrical potential $V_{Gr}$ is related to its charge density $\rho_{Gr}$ due to the graphene quantum capacitance: $V_{Gr} = V_{Gr}(\rho_{Gr})$, as described below.

The graphene surface electrical potential $V_{Gr}(\rho_{Gr})$, namely its vacuum level, can be determined in the following way. Since the graphene is connected to a voltage source, meaning its chemical potential is fixed externally, then the change of the graphene surface electrical potential $\Delta V_{Gr} = V_{Gr} - V_{Gr0}$ can be determined as $\Delta V_{Gr} = \Delta E_f/e$, where $\Delta E_f = E_f - E_{f0}$ is the graphene Fermi level change. Here $V_{Gr0}$ and $E_{f0}$ are the graphene surface electrical potential and Fermi level at charge neutrality ($\rho_{Gr} = 0$). For simplicity, we assume $V_{Gr0} = 0$ and $E_{f0} = 0$. Therefore, we have $V_{Gr} = E_f/e$. The relation between the graphene Fermi level $E_f$ and its carrier density $n_{Gr}$ is determined by

$$n_{Gr} = \begin{cases} \int_0^{E_f} DOS_{Gr}(E) dE, & (E_f \geq 0) \\ \int_{E_f}^{0} DOS_{Gr}(E) dE, & (E_f < 0) \end{cases},$$

where $DOS_{Gr}(E) = \frac{2}{\pi} \cdot \frac{|E|}{(\hbar v_F)^2}$ is the graphene density of states per unit area. Here $v_F = 10^6 m/s$ is the Femi velocity of the graphene. Using $\rho_{Gr} = -en_{Gr}$, we obtain the following boundary condition for the graphene plane

$$V_{Gr}(\rho_{Gr}) = \begin{cases} \frac{\sqrt{\pi}\hbar v_F}{e} \cdot \sqrt{-\frac{\rho_{Gr}}{e}}, & (\rho_{Gr} < 0) \\ -\frac{\sqrt{\pi}\hbar v_F}{e} \cdot \sqrt{\frac{\rho_{Gr}}{e}}, & (\rho_{Gr} \geq 0) \end{cases},$$

This boundary condition reflects the quantum capacitance of the graphene and its partial screening effect.

In the simulation, the boundary conditions for the tip and backgate are set at $V = V_{bias}$ and $V = V_{BG}$, respectively. To the make the simulation more computable we have a cylindrically truncated grounded surface enclose the simulation center. The dielectric constants in the region above and below the graphene are set to $\varepsilon_{vac} = \varepsilon_0$ and $\varepsilon_{hBN} = 4.2\varepsilon_0$, respectively, where $\varepsilon_0$ is the vacuum dielectric constant.

We note that although the tip height and tip cone angle are two independent parameters, they work together to control the value of $\alpha$ for the charged site. Therefore, in the simulation we fix one parameter and tune the other parameter to fit the experiment results. Here we fix the tip height at $h = 1nm$ and vary the value of $\theta$. This selected tip height is close to the STM tip tunneling distance used in the experiment. However, we note that the selection of $h$ here does not significantly affect the final obtained values for $\alpha$ and $\beta$, as discussed later. In the simulation $\alpha$ and $\beta$ are obtained via monitoring the responses of the electrical potential change $\Delta\Phi$ at the position of the charged site ($r = 5.4nm$ when the tip is fixed at the three-site symmetric point) in the TMD layer with the following parameters: $\alpha$: $V_{bias}$ = 145mV and $V_{BG}$ = 0, and $\beta$: $V_{bias}$ = 0 and $V_{BG}$ = 71 mV. The simulated electrical potential map at $V_{bias}$ = 0 and $V_{BG}$ = 71mV is shown in Fig. S3b while the simulated electrical potential map at $V_{bias}$ = 145mV and $V_{BG}$ = 0 is shown in Fig. S3c.

The simulated ratio between $\alpha$ and $\beta$ can be compared with the experimentally measured charging peak slope $\frac{\alpha}{\beta} = 0.507$. The parameters that yield results most consistent with experimental results are as follows: $\theta = 54.95°$, and $h = 1nm$ (Fig. S3d). The simulated tip coupling constant ($\alpha$) and backgate coupling constant ($\beta$) are $\alpha = 0.16$ and $\beta = 0.32$ for the charged site in the TMD layer ($r = 5.4nm$). We note that this simulated coupling constant is close to values reported in previous work[1].

The potential response of the charged site $\Delta\Phi$ is not exactly linear with the change of $V_{bias}$ and $V_{BG}$. This is shown in Fig. S3e ($\Delta\Phi$ as a function of $V_{bias}$) and S3f ($\Delta\Phi$ as a function of $V_{BG}$). This effect occurs because at large tip bias or backgate voltage the graphene is strongly doped and has a larger density of states at the Fermi level and behaves more like a metal. However, in the measurement range $V_{bias}$ < 200mV and $V_{BG}$ <100mV the response can still be approximately regarded as being linear.

The selection of the tip height does not significantly affect the final value for $\alpha$ in the simulation process above. This is seen in Fig. S3g and S3h, which show the simulated tip cone half angle $\theta$ (Fig. S3g) and the corresponding $\alpha$ values (Fig. S3h) for different tip heights. Including the uncertainty in the top hBN thickness and the hBN dielectric constant, we have an estimated value of $\alpha = 0.16 \pm 0.02$.

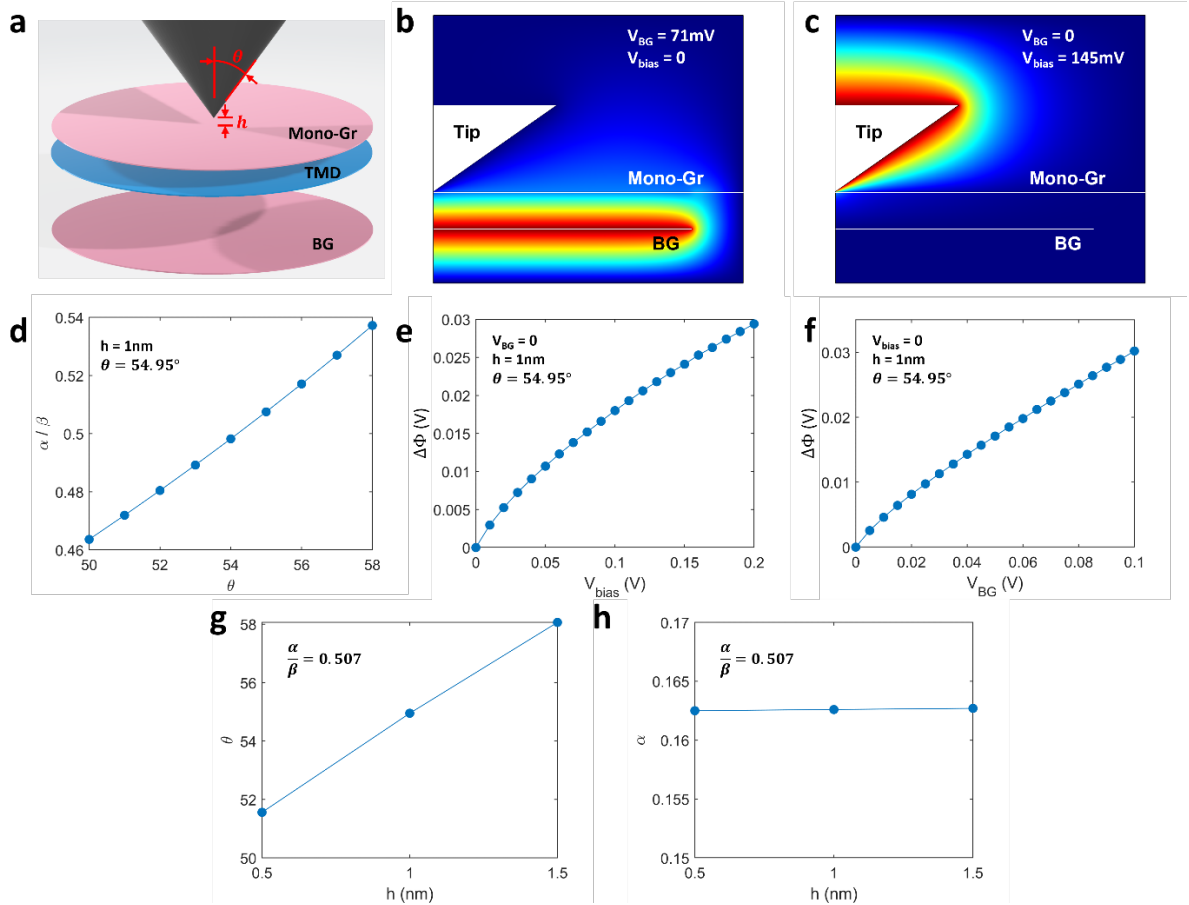

**Figure S3. Determination of the tip-TMD coupling constant. a.** Schematic of the simulation model. The tip is represented by an ideal metallic cone with half cone angle $\theta$ and tip height $h$ (separation between tip apex and the graphene surface). The backgate is modeled by an ideal metallic plate. The graphene is modeled as follow: We set the boundary conditions on the graphene surface to be that the electrical potential is determined by the charge density so that the quantum capacitance of the graphene can be correctly treated. See more details in section 3. **b-c.** Simulated electrical potential with (**b**) $V_{BG} = 71\text{mV}$ and $V_{bias} = 0$, and (**c**) $V_{BG} = 0$ and $V_{bias} = 145\text{mV}$. The simulation parameters used are $\theta = 54.95°$ and $h = 1nm$. The position of the t-WS$_2$ layer is not depicted here since it is regarded as an insulator with a dielectric constant equivalent to that of hBN. **d.** Simulated ratio $\alpha/\beta$ as a function of $\theta$ at h = 1nm. Here $\alpha$ ($\beta$) is obtained through monitoring the potential change $\Delta\Phi$ at r=5.4nm in the TMD layer by setting $V_{bias}$ = 145mV and $V_{BG}$ = 0 ($V_{bias}$ = 0 and $V_{BG}$ = 71mV). **e**. Potential change $\Delta\Phi$ at r=5.4nm in the TMD layer for different $V_{bias}$ values at $V_{BG}$ = 0. **f**. Potential change $\Delta\Phi$ at r=5.4nm in the TMD layer for different $V_{BG}$ values at $V_{bias}$ = 0. h = 1nm and $\theta = 54.95°$ for both (**e**) and (**f**). **g**. Fitted values for $\theta$ as a function of the tip height h with the ratio $\alpha/\beta$ fixed at 0.507. **g**. Fitted values for $\alpha$ as a function of the tip height h with the ratio $\alpha/\beta$ fixed at 0.507. The fitted value for $\alpha$ is nearly independent of the selected tip height h.